\documentclass[pdflatex,sn-mathphys-num]{sn-jnl}

\usepackage[T1]{fontenc}
\usepackage{graphicx}%
\usepackage{multirow}%
\usepackage{amsmath,amssymb,amsfonts}%
\usepackage{amsthm}%
\usepackage{mathrsfs}%
\usepackage[title]{appendix}%
\usepackage{xcolor}%
\usepackage{textcomp}%
\usepackage{manyfoot}%
\usepackage{booktabs}%
\usepackage{algorithm}%
\usepackage{algorithmicx}%
\usepackage{algpseudocode}%
\usepackage{listings}
\usepackage{slashed}%

\usepackage{bbm}
\usepackage{xcolor}
\usepackage[normalem]{ulem}

\newcommand{\LF}{\left(}
\newcommand{\RF}{\right)}
\newcommand{\LT}{\left[}
\newcommand{\RT}{\right]}

\newcommand{\Pc}{\mathcal{P}}
\newcommand{\Tc}{\mathcal{T}}
\newcommand{\Hc}{\mathcal{H}}
\newcommand{\Fc}{\mathcal{F}}
\newcommand{\Oc}{\mathcal{O}}
\newcommand{\Lc}{\mathcal{L}}

\newcommand{\Cc}{\mathcal{C}}

\begin{document}

\title[Article Title]{Hawking radiation with pure states}


\author*[1]{\fnm{} \sur{K. Sravan Kumar}}\email{sravan.kumar@port.ac.uk}

\author[2]{\fnm{Jo\~ao Marto} }\email{jmarto@ubi.pt}
\equalcont{These authors contributed equally to this work.}

\equalcont{These authors contributed equally to this work.}

\affil*[1]{\orgdiv{Institute of Cosmology and Gravitation}, \orgname{U. Portsmouth}, \orgaddress{\street{Dennis Sciama Building, Burnaby Road, United Kingdom}, \city{Portsmouth}, \postcode{PO1 3FX}, \country{United Kingdom}}}

\affil[2]{\orgdiv{Departamento de F\'isica, Centro de Matem\'atica e Aplica\c{c}\~oes (CMA-UBI)}, \orgname{Universidade da Beira Interior}, \orgaddress{\street{Rua Marqu\^es D \'Avila e Bolama}, \city{Covilh\~a}, \postcode{6200-001}, \country{Portugal}}}


\abstract{
Hawking's seminal work on black hole radiation highlights a critical issue in our understanding of quantum field theory in curved spacetime (QFTCS), specifically the problem of unitarity loss (where pure states evolve into mixed states). In this paper, we examine a recent proposal for a direct-sum QFTCS, which maintains unitarity through a novel quantization method that employs geometric superselection rules based on discrete spacetime transformations. This approach describes a quantum state in terms of components that evolve within geometric superselection sectors of the complete Hilbert space, adhering to the discrete symmetries of a Schwarzschild black hole. Consequently, it represents a maximally entangled pure state as a direct-sum of two components in the interior and exterior regions of the black hole, thereby preserving the unitarity of Hawking radiation by keeping it in the form of pure states. }




\maketitle

\section{Introduction}
\label{sec1}

The seminal papers of Hawking's on black hole (BH) radiation \cite{Hawking:1974rv,Hawking:1975vcx} have exposed explicitly the incompatibility in our understanding of gravity and quantum mechanics. To be precise, the issue of understanding the behavior of fields in black hole spacetime was first discussed by Einstein and Rosen in 1935 \cite{Einstein:1935tc}, even before the formal development of quantum field theory (QFT). The important conundrum is the loss of unitarity, which involves the evolution of pure states into mixed states. In other words, Hawking radiation is formed by mixed states, and the (maximally) entangled partner is always left behind the Schwarzschild black hole (SBH) horizon \cite{Mathur:2008nj,Calmet:2022swf}. This would mean an observer outside the SBH would access states whose fate depends on what happens to the partner states behind the horizon. This opens a serious conceptual conundrum: the observer's quantum physics depends on what is happening in the region of spacetime to which the observer has no causal access. Since an asymptotic observer would never witness any particle (quantum state) crossing the horizon, the violation of unitarity results in an incomplete understanding of quantum physics for the asymptotic observer. Often, it is interpreted as an acceptable outcome to treat black holes as special objects in the Universe that would violate unitarity \cite{Wald:1995yp}. But this interpretation is not the solution \cite{Kiefer_2001}, and even if it is, one is left with the prediction, by D. Page, that the entanglement entropy of Hawking quanta would at some point exceed the thermodynamic entropy of the black hole \cite{Page:1993wv,Page:2013dx,Buoninfante:2021ijy}. Often (Planck scale) quantum gravity proponents take solace in an unknown ultraviolet (UV) complete theory to resolve the problem of unitarity in black hole physics \cite{Almheiri:2020cfm,Haggard:2014rza}. This raises the logical question of why Planck-scale quantum gravity is necessary to address a fundamental problem that arises from understanding quantum fields in curved spacetime.
Along with unitarity loss, there is also a related issue of information loss, which carries an additional question on the universal nature of Hawking radiation that leaves no knowledge of what has formed the black hole. Understanding the formation of a black hole (quantum mechanically) is another hard question that requires a robust formulation of QFT in a dynamically collapsing geometry, which we do not have yet. Hawking's actual calculation involves quantization in Schwarzschild spacetime, even though he framed the discussion in the context of a black hole formed by collapse. Despite numerous attempts \cite{Raju:2020smc,Raju:2021lwh} to address the information paradox issue, Einstein-Rosen's initial question of particle problem in general relativity remains and needs to be taken very seriously.
Recent ongoing efforts by Gerard 't Hooft\cite{tHooft:2015pce,tHooft:2016qoo,tHooft:2016rrl,tHooft:2016sdu,tHooft:2018waj,tHooft:2021itz,tHooft:2021lyt,tHooft:2022bgo}, building on the work of Norma G. Sanchez and B. F. Whiting \cite{Sanchez:1986qn}  and the gravitational backreaction calculations by Dray and 't Hooft \cite{Dray:1984ha}, underscore the importance of resolving the information paradox through fundamental physics before turning to an unknown theory of quantum gravity.

The recently proposed direct-sum quantum field theory in curved spacetime (QFTCS) presents an elegant new approach to addressing the issues of unitarity and information loss in black hole physics \cite{Kumar:2023ctp,Kumar:2023hbj,Kumar:2024oxf}. The direct-sum QFTCS has also led to the promising explanation for the long-standing CMB anomalies in the context of inflationary cosmology \cite{Gaztanaga:2024vtr,Gaztanaga:2024whs}. The framework of direct-sum QFTCS is based on the intricate understanding of discrete spacetime symmetries and separating the notions of classical and quantum mechanical time. In this review, we explore how this framework provides a unitary description of Hawking radiation and offers significant insights into extracting information about the black hole interior through a fundamental approach. Black holes and de Sitter spacetime share certain analogies related to event horizons, as explored by Gibbons and Hawking \cite{Gibbons:1977mu}. Thus, we briefly address a similar unitarity issue in de Sitter space and the potential resolution offered by direct-sum QFTCS.

This review is organized as follows: in Sec.~\ref{sec:disumQM}, we discuss the parity ($\mathcal{P}$) and time reversal ($\mathcal{T}$) symmetries in quantum mechanics and Minkowski QFT. We then present the foundational framework of direct-sum QFT. In Sec.~\ref{sec:Hawking}, we discuss the Schwarzchild BH spacetime and the basic assumptions in Hawking's original calculation. We then present the details of 't Hooft's quantum algebra that emerges from the effects of gravitational backreaction. In Sec.~\ref{sec:disumBH}, we summarize the results of direct-sum QFT in BH spacetime and its approach to obtaining Hawking radiation in the form of pure states. We also show the QFT extension of 't Hooft quantum algebra and the resulting guidelines to extract information from Hawking radiation.  In Sec.~\ref{sec:con}, we conclude with a qualitative discussion and present outlook for future investigations. 

Throughout the manuscript, we use the metric signature mostly positive $\LF -,\,+,\,+,\,+ \RF$ and $\hbar = c=1$. 

\section{$\mathcal{P}\mathcal{T}$ symmetry and direct-sum quantum theory}

\label{sec:disumQM} 

Quantum mechanics is recognized for its time symmetry \cite{tHooft:2018jeq, Donoghue:2019ecz}, a property linked to the anti-unitary nature of the time reflection operation. In quantum theory, time is treated as a parameter rather than an operator, unlike spatial position. This distinct treatment of time sets it apart in quantum theory compared to classical theory. Quantum Field Theory (QFT) in Minkowski spacetime represents a unification of special relativity and quantum mechanics, achieved by enforcing the commutation of field operators at space-like separations.
In the context of Klein-Gordon field this is expressed as
\begin{equation}
    \LT \hat \phi(x),\,\hat \phi(y)\RT = 0,\quad \LF x-y\RF^2>0\,. 
    \label{eq:caus}
\end{equation}
The above condition together with Lorentz invariance dictates the expansion of Klien-Gordon (KG) (real scalar) field operator as  
\begin{equation}
    \hat{\phi}(x) = \int \frac{d^3k}{\LF 2\pi \RF^{3/2}}\frac{1}{\sqrt{2\vert k_0\vert }}\LT a_{\textbf{k}}e^{ik\cdot x}+a_{\textbf{k}}^\dagger e^{-ik\cdot x}   \RT 
    \label{eq:kgop}
\end{equation}
where $k\cdot x= -k_0t+\textbf{k}\cdot \textbf{x}$. The connection with quantum mechanics stems from the positive energy state definition, defined by 
\begin{equation}
    \vert \Psi\rangle_t = e^{-iEt}\vert \Psi\rangle_0
    \label{eq:posE}
\end{equation}
assuming the arrow of time to be $t: -\infty \to \infty$. We can see that the expression of KG field operator \eqref{eq:kgop} is a combination of positive and negative energy solutions. These are associated with creation and annihilation operators $\LF a_\textbf{k},\, a_\textbf{k}^\dagger  \RF$ which satisfy the canonical commutation relations and the Minkowski vacuum is defined by 
\begin{equation}
    \LT a_{\textbf{k}},\, a_{\textbf{k}}^\dagger  \RT = 1,\quad a_\textbf{k}\vert 0_M\rangle =0.
\end{equation}
Another equivalent way of defining a positive energy state \eqref{eq:posE} is 
\begin{equation}
    \vert \Psi\rangle_t = e^{iEt}\vert \Psi\rangle_0
    \label{altdef}
\end{equation}
with the presumption on the arrow of time $t: \infty \to -\infty$. Usually, the entire quantum theory is constructed based on the convention \eqref{eq:posE}. As noted in \cite{Donoghue:2020mdd}, one can redefine the entire framework of quantum theory by changing the definition of positive energy \eqref{altdef}, which means reversing the convention on the arrow of time and effectively replacing $i$ with $-i$ throughout the quantum (field) theory.

Direct-sum quantum theory starts with a construction that combines both arrows of time together with parity operation to express a single quantum state. According to this, a single quantum state is expressed by joining two components through a direct-sum operation 
\begin{equation}
    \vert \Psi\rangle = \frac{1}{\sqrt{2}} \LF \ \vert\Psi_+\rangle \oplus \vert \Psi_-\rangle \RF = \frac{1}{\sqrt{2}}\begin{pmatrix}
        \vert \Psi_+\rangle \\ 
        \vert \Psi_-\rangle 
    \end{pmatrix}
    \label{dss}
\end{equation}
 $\vert\Psi_{\pm}\rangle $ are orthogonal states  \cite{Conway} of the direct-sum Hilbert space 
 \begin{equation}
     \Hc = \Hc_+ \oplus \Hc_-
 \end{equation}
Here, $\Hc_\pm$ are called the geometric superselection sectors\footnote{The nomenclature geometric superselection sectors is because the Hilbert spaces $\Hc_\pm$ are associated with the physical states in the parity conjugate regions, which means we attach the Hilbert space to describe states in a particular region of physical space.}; states in these sectors cannot be superposed coherently with each other, and a state corresponding to one sector does not evolve to the other. Here $\pm$ mean the parity conjugate regions of physical space. The states $\vert\Psi_{\pm}\rangle $ evolve according the following direct-sum Schr\"{o}dinger equation 
\begin{equation}
    i\frac{\partial\vert \Psi\rangle}{\partial t_p} = \begin{pmatrix}
			\hat{{H}}_+ & 0 \\ 
			0 & -\hat{{H}}_-
		\end{pmatrix} \vert \Psi\rangle
  \label{disumSch}
\end{equation}
where $\hat{H}_\pm\LF \hat{x}_\pm,\,\hat{p}_\pm \RF$ give the Hamiltonians of the system $\hat{H}=\hat{H}_+\oplus \hat{H}_-$, which are functions of position and momenta operators of the target space defined by
\begin{equation}
\begin{aligned}
\hat{p}_+ & = -i\frac{d}{dx_+},\quad x_+=x\gtrsim 0 \\ 
\hat{p}_- & = i\frac{d}{dx_-},\quad x_-=x\lesssim 0\,. 
\end{aligned}
\end{equation}
{where the eigenvalues of $\hat x_{\pm}$ are parity conjugate points which mean $x_+\in (0,\,\infty]$ and $x_-\in [-\infty,\,0)$. }

The canonical non-zero commutation relations are\footnote{{The remaining relations are $\LT \hat x_+,\, \hat x_- \RT = \LT \hat p_+,\, \hat p_- \RT = \LT \hat x_+,\, \hat p_- \RT = \LT \hat p_+,\, \hat x_- \RT =0$.} } 
\begin{equation}
    [\hat{x}_+,\,\hat{p}_+] = i,\quad [\hat{x}_-,\,\hat{p}_-]=-i\,,
\end{equation}
The complete description of a quantum state, in the entire physical space, is given by $\vert \Psi\rangle $
\begin{equation}
\Psi(x) = \frac{1}{\sqrt{2}} \begin{pmatrix}
       \langle x_+\vert \quad \langle x_- \vert 
    \end{pmatrix}\begin{pmatrix}
        \vert \Psi_+\rangle_0 e^{-iEt} \\
        \vert\Psi_-\rangle_0 e^{iEt}  
    \end{pmatrix} \implies \begin{cases}
        \frac{1}{\sqrt{2}}\Psi_+\LF x_+ \RF e^{-iEt},\quad x_+ = x\gtrsim 0 \\  \frac{1}{\sqrt{2}}\Psi_-\LF x_- \RF e^{iEt},\quad x_- = x\lesssim 0\,.
    \end{cases}
    \label{eq:wavefunction}
\end{equation}

The square integrability and the probabilities of states are given by 
\begin{equation}
  \int_{-\infty}^\infty dx  \langle \Psi\vert \Psi \rangle =  \frac{1}{2} \int^{0}_{-\infty} dx_-  \langle \Psi_-\vert \Psi_- \rangle +\frac{1}{2} \int^{\infty}_0 dx_+  \langle \Psi_+\vert \Psi_+ \rangle   =1\,. 
\end{equation}
\begin{figure}
     \centering
     \includegraphics[width=0.5\linewidth]{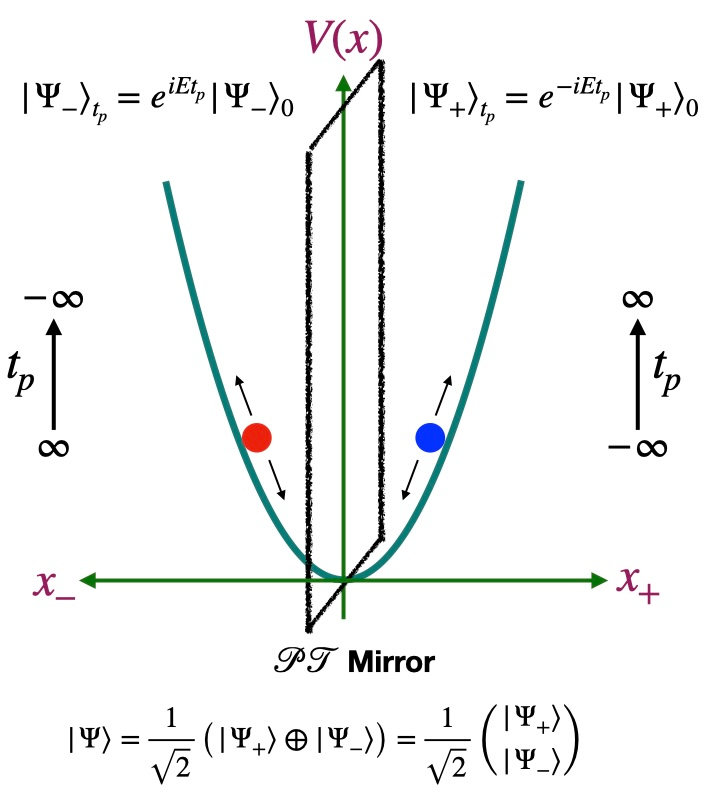}
     \caption{This is a pictorial description of how a quantum harmonic oscillator is described by direct-sum quantum theory by means of two quantum states, evolving with two opposite arrows of time at parity conjugate points in physical space. }
     \label{fig:disumosc}
 \end{figure}
The description of a harmonic oscillator, in the current approach, can be hinted from Fig.~\ref{fig:disumosc}. The wave function of the parity conjugate region is described by 
\begin{equation}
   \Psi_n(x) = \langle x \vert \Psi_n\rangle \equiv \begin{cases}
       \frac{1}{\sqrt{2^{n+1}n!}}\LF \frac{1}{\pi} \RF^{1/4} e^{-\frac{1}{2}x_+^2} H_n\LF x_+ \RF e^{-iE_n t_p},\quad x_+\gtrsim 0\\ 
      \frac{1}{\sqrt{2^{n+1}n!}}\LF \frac{1}{\pi} \RF^{1/4} e^{-\frac{1}{2}x_-^2} H_n\LF x_- \RF e^{iE_n t_p},\quad x_- \lesssim 0
   \end{cases}
\end{equation}
where $E_n$ corresponds to energy state for each $n$ of the Hermite polynomials $H_n$. We can see that in the limit $x_\pm \to 0_\pm$, the two-component states $\vert \Psi_{\pm}\rangle$ match automatically. The point $x=0$ is not a special point since under the translation $x\to x+L$, the construction remains the same because it is based on $\Pc\Tc$, a discrete transformation. The orthogonality of states corresponding to different energy levels is given by
 \begin{equation}
     \int_{-\infty}^{\infty} \langle \Psi_n\vert \Psi_m\rangle dx = \int_0^\infty dx_+ \langle \Psi_{n+}\vert \Psi_{m+}\rangle + \int_{-\infty}^0 dx_- \langle \Psi_{n-}\vert \Psi_{m-}\rangle dx_- = \delta_{n,m}
 \end{equation}
Furthermore, an observer can only measure the quantum state in either of the parity conjugate points, and the arrow of time in quantum theory is not observable for any classical observer. 

Minkowski spacetime $ds^2= -dt^2+d\textbf{x}^2$ is $\Pc\Tc$ symmetric (i.e., metric is invariant under $t\to-t$ and $\textbf{x}\to -\textbf{x}$). Thus, one can extend the direct-sum Schr\"{o}dinger equation \eqref{dss} to the relativistic case by direct-sum QFT where a KG field operator is expressed as direct-sum of two components as a function of $\Pc\Tc$ conjugate points 
\begin{equation}
    \hat \phi(x)  = \frac{1}{\sqrt{2}}\begin{pmatrix}
        \hat \phi_+ & 0 \\ 
        0 & \hat \phi_-
    \end{pmatrix}
    \label{eq:qfdisum}
\end{equation}
where 
\begin{equation}
    \begin{aligned}
    \hat \phi_{+}( x) & =     \int \frac{d^3k}{\LF 2\pi \RF^{3/2}}\frac{1}{\sqrt{2\vert k_0\vert }}\LT a_{(+)\textbf{k}}e^{ik\cdot  x}+a_{(+)\textbf{k}}^\dagger e^{-ik\cdot x}   \RT \\ 
    \hat \phi_-(- x) & =    \int \frac{d^3k}{\LF 2\pi \RF^{3/2}}\frac{1}{\sqrt{2\vert k_0\vert }}\LT a_{(-)\textbf{k}}e^{-ik\cdot x}+a_{(-)\textbf{k}}^\dagger e^{ik\cdot  x}\RT 
    \end{aligned}
    \label{eq:fppm}
\end{equation}
The creation and annihilation operators obey 
\begin{equation}
    \LT a_{(\pm)\textbf{k}},\, a^\dagger_{(\pm)\textbf{k}}\RT = 1,\quad  \LT a_{(\pm)\textbf{k}},\, a^\dagger_{(\mp)\textbf{k}}\RT= \LT a_{(\pm)\textbf{k}},\, a_{(\mp)\textbf{k}}\RT =0\,.
\end{equation}
which imply a new causality condition 
\begin{equation}
    \LT \hat \phi_+(x),\, \hat \phi_-(-y) \RT =0\,.
\end{equation}
Here we notice that, in our notation, $\hat \phi_{\pm}$ are field operators defined for parity conjugate points in physical space, which mean the first line in \eqref{eq:fppm} is defined only for $\textbf{x}\gtrsim 0 $ where as the second line is defined only for $\textbf{x}\lesssim 0 $, the direct-sum of these two operators define the quantum field \eqref{eq:qfdisum} everywhere in Minkowski spacetime. Since the construction is based on $\Pc\Tc$, any Lorentz transformation on \eqref{eq:qfdisum} preserve the structure of $\Pc\Tc$ symmetric Minkowski vacuum given by
\begin{equation}
    \vert 0_M \rangle = \frac{1}{\sqrt{2}}\begin{pmatrix}
        \vert 0_{M+}\rangle \\ 
        \vert 0_{M-}\rangle 
    \end{pmatrix},\quad a_{(+)\textbf{k}}\vert 0_{M+}\rangle =0,\quad a_{(-)\textbf{k}}\vert 0_{M-}\rangle =0\,. 
\end{equation}
The two-point function and propagator in direct-sum QFT are given by
\begin{equation}
\begin{aligned}
    \langle 0\vert \hat{\phi}\LF x \RF  \hat{\phi}\LF x^\prime \RF \vert 0\rangle & = \frac{1}{2} \langle 0_+\vert \hat{\phi}_+\LF x \RF  \hat{\phi}_+\LF x^\prime \RF \vert 0_+\rangle + \frac{1}{2} \langle 0_-\vert \hat{\phi}_-\LF -x \RF  \hat{\phi}_-\LF -x^\prime \RF \vert 0_- \rangle \\ 
     \langle 0\vert T \hat{\phi}\LF x \RF  \hat{\phi}\LF x^\prime \RF 0\vert \rangle & = \frac{1}{2} \langle 0_+\vert T\hat{\phi}_+\LF x \RF  \hat{\phi}_+\LF x^\prime \RF \vert 0_+\rangle + \frac{1}{2} \langle 0_-\vert T \hat{\phi}_-\LF -x \RF  \hat{\phi}_-\LF -x^\prime \RF \vert 0_- \rangle \,, 
    \end{aligned}
\end{equation}
where $T$ represents time ordering. All the interactions are divided into direct-sum components. For example, cubic interaction like the following
\begin{equation}
  \frac{\lambda}{3} \hat{\phi}^3 = \frac{\lambda}{3} \begin{pmatrix}
      \hat{\phi}_+^3 & 0 \\ 
      0 & \hat{\phi}_-^3
  \end{pmatrix}
\end{equation}
shows that there will never be any mixing between $\hat{\phi}_+$ and $\hat{\phi}_-$. Therefore, all the standard QFT calculations can be straightforwardly extended to DQFT but we do not see any change in results because Minkowski spacetime is $\Pc\Tc$ symmetric (See \cite{Kumar:2023ctp,Kumar:2024oxf,GKM} for more details). According to the DQFT standard model, vacuum and the degrees of freedom of particles ($\vert SM\rangle$) and antiparticles ($\vert \overline{SM}\rangle$ are represented by
\begin{equation}
    \vert 0_{SM}\rangle = \begin{pmatrix}
        \vert 0_{SM+}\rangle \\ 
        \vert 0_{SM-}\rangle 
    \end{pmatrix} \quad \vert SM\rangle = \frac{1}{\sqrt{2}}\begin{pmatrix}
        \vert SM_+\rangle \\ 
        \vert SM_-\rangle \end{pmatrix} \quad \vert \overline{SM}\rangle = \frac{1}{\sqrt{2}}\begin{pmatrix}
        \vert \overline{SM}_+\rangle \\ 
        \vert \overline{SM}_-\rangle 
    \end{pmatrix}
\end{equation}
{Note that we apply the same geometric super-selection-rule based on $\Pc\Tc$ for all Fock spaces of the standard model degrees of freedom, i.e., the parity conjugate regions are uniquely defined for all the particle and antiparticle states of the standard model. Note that when we include charge conjugation operation $\Cc$, every quantum field is again expressed as a direct-sum of components in geometric superselection sectors of Fock space defined by $\Pc\Tc$. The standard understanding of $\Cc\Pc\Tc$ symmetry in QFT holds separately in both the geometric superselection sectors, as further explained below. 

 While this work primarily addressed the DQFT quantization of a real scalar field, the approach naturally extends to complex scalars, fermions, and gauge fields \cite{GKM}. In the DQFT framework, each quantum field is expressed as a direct-sum of two components, acting as $\Pc\Tc$ mirror reflections of each other, spanning the full Minkowski spacetime. Consequently, the conventional methods of field quantization can be easily adapted to fit within the DQFT structure.

\begin{itemize}
    \item Complex scalar field operator $\hat{\phi}_c  = \frac{1}{\sqrt{2}}\LF \hat{\phi}_{c\,+}\oplus  \hat{\phi}_{c\,-}\RF$ in DQFT is expanded as 
   \begin{equation}
   \begin{aligned}
    &  \hat{\phi}_{c\,\pm} = \int  \frac{d^3k}{\LF 2\pi \RF^{3/2}}\frac{1}{\sqrt{2\vert k_0\vert }}\Bigg[ a_{(\pm)\textbf{k}}e^{\pm ik\cdot  x}+b_{(\pm)\textbf{k}}^\dagger e^{\mp ik\cdot x}   \Bigg] \\ & \LT \hat{\phi}_{c\,+},\, \hat{\phi}_{c\,-} \RT =0\,,
    \end{aligned}
   \end{equation}
   where $a_{(\pm)\textbf{k}},\, a^\dagger_{(\pm)\textbf{k}}$ and $b_{(\pm)\textbf{k}},\, b^\dagger_{(\pm)\textbf{k}}$ are canonical creation and annihilation operators of the parity conjugate regions (denoted by subscripts $_{(\pm)}$) attached with geometric superselection sector.
   All the cross commutation relations of $a_{(\pm)},\, a^\dagger_{(\pm)}$ and $b_{(\pm)},\, b^\dagger_{(\pm)}$ vanish.  
   \item Fermionic field operator $ \hat \psi = \frac{1}{\sqrt{2}}\LF \hat \psi_+\oplus \hat \psi_- \RF$ in DQFT becomes 
   \begin{equation}
     \hat  \psi_{\pm} = \sum_{{\tilde s}} \int \frac{d^3k}{\LF 2\pi \RF^{3/2}\sqrt{2\vert k_0\vert}} \Bigg[ c_{{\tilde s}(\pm)\textbf{k}} u_{\tilde s}(\textbf{k}) e^{\pm ik\cdot x} + d_{{\tilde s}(\pm)\textbf{k}}^\dagger v_{\tilde s}(\textbf{k}) e^{\mp ik\cdot x}\Bigg]
   \end{equation}
where ${\tilde s}=1,2$ correspond to the two independent solutions of $\LF \slashed k+m\RF u_s=0$ and $\LF -\slashed k+m\RF v_s=0$ corresponding to spin-$\pm\frac{1}{2}$. The creation and annihilation operators, of the Fock space geometric superselection sector, satisfy the anti-commutation relations $\Big\{ c_{s(\pm)\textbf{k}},\,c_{s(\pm)\textbf{k}}^\dagger \Big\}=1,\, \Big\{ c_{s(\mp)\textbf{k}},\,c_{s(\pm)\textbf{k}}^\dagger \Big\}=\Big\{ c_{s(\mp)\textbf{k}},\,c_{s(\pm)\textbf{k}} \Big\}=0$ leading to the new causality condition $\Big\{ \hat\psi_+,\,\hat \psi_-\Big\} =0$.
\item The vector field operator $\hat A_\mu = \frac{1}{\sqrt{2}}\LF \hat{A}_{+\mu}\oplus \hat A_{-\mu} \RF$ in DQFT expressed as 
\begin{equation}
	\hat{A}_{\pm \mu}= \int \frac{d^3k}{\LF 2\pi \RF^{3/2}\sqrt{2\vert k_0\vert }} e^{(\lambda)}_\mu\Bigg[ c_{(\pm \lambda)\textbf{k}} e^{\pm ik\cdot x}+c^\dagger_{(\pm \lambda)\textbf{k}} e^{\mp ik\cdot x}  \Bigg]
	\end{equation} 
	where $e^{(\lambda)}_\mu$ is the polarization vector satisfying the transverse and traceless conditions. The creation and annihilation operators $c_{(\pm \lambda)\textbf{k}},\,c_{(\pm \lambda)\textbf{k}}^\dagger$ satisfy the similar relations as \eqref{eq:fppm}.
\end{itemize}

All the Standard model calculations remain the same because all the interactions terms become 
\begin{equation}
   \Lc_c \sim\Oc_{SM}^3=\begin{pmatrix}
        \Oc_{SM_+}^3 & 0 \\ 
        0 & \Oc_{SM_-}^3
    \end{pmatrix} \quad \Lc_q \sim \Oc_{SM}^4 = \begin{pmatrix}
        \Oc_{SM_+}^4 & 0 \\ 
        0 & \Oc_{SM_-}^4
    \end{pmatrix}
\end{equation}
where $\Oc_{SM}$ is any operator in the Standard Model involving quantum fields and their derivatives. }Therefore, direct-sum QFT is a framework that does not alter the QFT calculations in Minkowski due to the spacetime being $\Pc\Tc$ symmetric.
For example, when calculating a scattering amplitude, such as the transition from N particles to M particles, the outcome under DQFT remains identical to that of standard QFT, namely
\begin{equation}
\begin{aligned}
    A_{N\to M} &= \frac{A^{N\to M}_+ \LF p_a, -p_b\RF  + A^{N\to M}_-\LF -p_a, p_b \RF}{2}\\
    A^{N\to M}_+ \LF p_a, -p_b\RF &= A^{N\to M}_-\LF -p_a, p_b \RF,
\end{aligned}
\end{equation}
where $p_a,\,p_b$ with $a=1,\cdots N$ and $b=1,\cdots M$ represent the 4-momenta of all the states involved in the scattering. 
$A_{\pm}$ represent amplitudes as a function of 4-momenta of initial and final states
computed in both vacuums $\vert 0_{SM\pm}\rangle$. Note that the in and out states in $\vert 0_{SM\pm}\rangle$ have opposite signs, a consequence of the reversed arrow of time in the two vacuums. The amplitudes $A_{\pm}$ remain equal at all orders of perturbation theory, as a result of the $\Pc\Tc$ symmetry inherent to Minkowski spacetime.
The well-known $CPT$ invariance (charge conjugation, parity, and time reversal) of scattering amplitudes \cite{Coleman:2018mew} remains valid in both vacuums as well, ensuring the symmetry is preserved across all states, which means
\begin{equation}
\begin{aligned}
    A^{N\to M}_+(p_a, -p_b) & = A^{M\to N}_+(-p_a, p_b) \\
    A^{N\to M}_-(-p_a, p_b) &= A^{M\to N}_-(p_a, -p_b) \,.
\end{aligned}
\end{equation}
This is attributed to the fact that the CPT operation of any scattering process would turn the outgoing anti-particles into in-going particles and vice-versa \cite{Coleman:2018mew}.  

However, this new understanding provides a fresh perspective on QFT in curved spacetime, ultimately leading to unitarity and observer complementarity in black hole spacetime, as proposed by Susskind \cite{Susskind:1993if}. What was previously only a hypothesis is now supported by our framework, offering a fundamental approach to achieving these principles.

\section{On the assumptions behind Hawking's calculation and 't Hooft Gravitational backreaction}

\label{sec:Hawking}

In this section, we discuss the crucial observations of Hawking calculation \cite{Hawking:1974rv,Hawking:1975vcx} that has lead to the unitarity loss and information paradox \cite{Kumar:2023hbj}. Hawking's calculation is concerned with the quantization of scalar field in Schwarzschild BH spacetime given by 
\begin{equation}
	ds^2 = -\LF 1-\frac{2GM}{r} \RF dt^2 + \frac{1}{\LF 1-\frac{2GM}{r} \RF}dr^2 + r^2d\Omega^2\,,
	\label{SBHmet}
\end{equation}
SBH spacetime is static and spherically symmetric and invariant under the discrete transformations 
\begin{equation}
	\Tc: 	t\to -t,\quad \Pc: \LF \theta,\,\phi \RF \to \LF \pi-\theta,\,\pi+\varphi \RF 
	\label{SBHdis}
\end{equation}
and asymptotically Minkowski i.e., $r\to \infty$. The coordinate singularity at $r=2GM$ makes the Schwarzschild coordinates unfit for understanding quantum fields (which was first found by Einstein-Rosen in 1935 \cite{Einstein:1935tc}). This is why Kruskal-Szekers (KS) coordinates are often used in the literature instead. In those coordinates, the SBH metric is given by
\begin{equation}
	ds^2 = - \frac{2GM}{r}e^{-\frac{r}{2GM}} dUdV + r^2d\Omega^2\,, 
	\label{UVmetric}
\end{equation}
where 
\begin{equation}
UV = 16G^2M^2 e^{\frac{r}{2GM}}\LF 1- \frac{r}{2GM} \RF 
	\label{UVcon}
\end{equation}
The relation between KS coordinates (U,\,V) and the Schwarzschild coordinates is given by 
\begin{equation}
	r>2GM \implies  \begin{cases}
		U= -4GM e^{-\frac{u}{4GM}} <0 & 	V= 4GM e^{\frac{v}{4GM}}>0  \\ U= 4GM e^{-\frac{u}{4GM}} >0 &	V= -4GM e^{\frac{v}{4GM}} <0\,.
	\end{cases}
	\label{rg2M}
\end{equation}
and 
\begin{equation}
	r<2GM \implies  \begin{cases}
		U= 4GM e^{-\frac{u}{4GM}} >0 & 	V= 4GM e^{\frac{v}{4GM}}>0  \\ U= -4GM e^{-\frac{u}{4GM}} <0 &	V= -4GM e^{\frac{v}{4GM}} <0\,.
	\end{cases}
	\label{rg2M2}
\end{equation}
where $u = t-r_\ast$ and $v=t+r_\ast$ with $r_\ast= r+2GM\ln \Bigg\vert \frac{r}{2GM} -1\Bigg\vert$ being the so-called tortoise coordinate which asymptote to $r_\ast \to -\infty $ as $r\to 2GM$. Thus
\begin{equation}
   r\to 2GM  \implies \begin{cases}
		U\to 0^{\mp} & V\to 0^{\pm} \, \text{for}\quad r>2GM \\ 
		U\to 0^{\pm} & V\to 0^{\pm} \, \text{for}\quad r<2GM 
	\end{cases}
\end{equation}
As \eqref{SBHmet} coincides with Minkowski spacetime in the limit $r\to \infty$,  and if one assumes an arrow of time $t: -\infty \to \infty$ for the quantum fields in this asymptotic Minkowski spacetime, we find that Hawking considers $U<0,\, V>0$ to be the only physical region in the exterior $r>2GM$. Thus, the other possibility $U>0,\, V<0$ is considered to be a parallel or an unphysical Universe. In a similar way, $U<0,\,V<0$ (often called a white hole) of $r<2GM$ is declared to be unphysical too. This is because the spacetime, long before the black hole has formed, is assumed to be Minkowski with an arrow of time $t: -\infty \to \infty$. Thus, the formation of spacetime with $U<0,\,V<0$ {and $U>0,\,V<0$} was thought to be an impossible scenario from a collapsing geometry. There are two caveats in these assumptions 
\begin{itemize}
    \item We must be very careful with taking classical intuitions into formulating the description of quantum fields in curved spacetime. Historically, quantum physics is known to be a counterintuitive formulation. 
    \item Understanding the formation of BH quantum mechanically is a profound open question; thus, declaring $U<0,\,V<0$ and $U>0,\,V<0$ as irrelevant for quantum fields in SBH spacetime has to be taken with great care. 
    \item Symmetries are a guiding concept in physics, and one must be careful in discarding them (by hand). Given that the following (discrete) spacetime transformations on SBH metric \eqref{UVcon}
    \begin{equation}
        \LF U,\,V \RF \to \LF -U,\, -V \RF,\quad \LF \theta,\,\varphi \RF\to \LF \pi-\theta,\,\pi+\varphi  \RF
        \label{dssym}
    \end{equation}
    leave the metric \eqref{UVmetric} invariant, there is an ambiguity on the choice of the sign of the coordinates, and it was pointed out by Einstein and Rosen as an important ordeal for combining gravity and quantum mechanics (See \cite{GKM}). 
\end{itemize}

Unitarity loss is the hypothetical result of restricting quantum fields to only certain regions of the KS coordinates $\LF U<0,\,V>0 \RF$ and $\LF U>0,\,V>0\RF$ whereas the other regions $\LF U>0,\,V<0 \RF$ and $\LF U<0,\,V<0\RF$ are equally allowed by symmetry. Indeed, the Hartle-Hawking vacuum, obtained from the gravitational path integral (in Euclidean time), gives the relation between the quantum states connecting the previously omitted regions  $\LF U>0,\,V<0 \RF$ and $\LF U<0,\,V<0\RF$ \cite{Hartle:1976tp}. It is not a surprising revelation because allowing a complex time coordinate $t\to it$ would render all arrows of time possible. 

\begin{figure}
	\centering
	\includegraphics[width=0.7\linewidth]{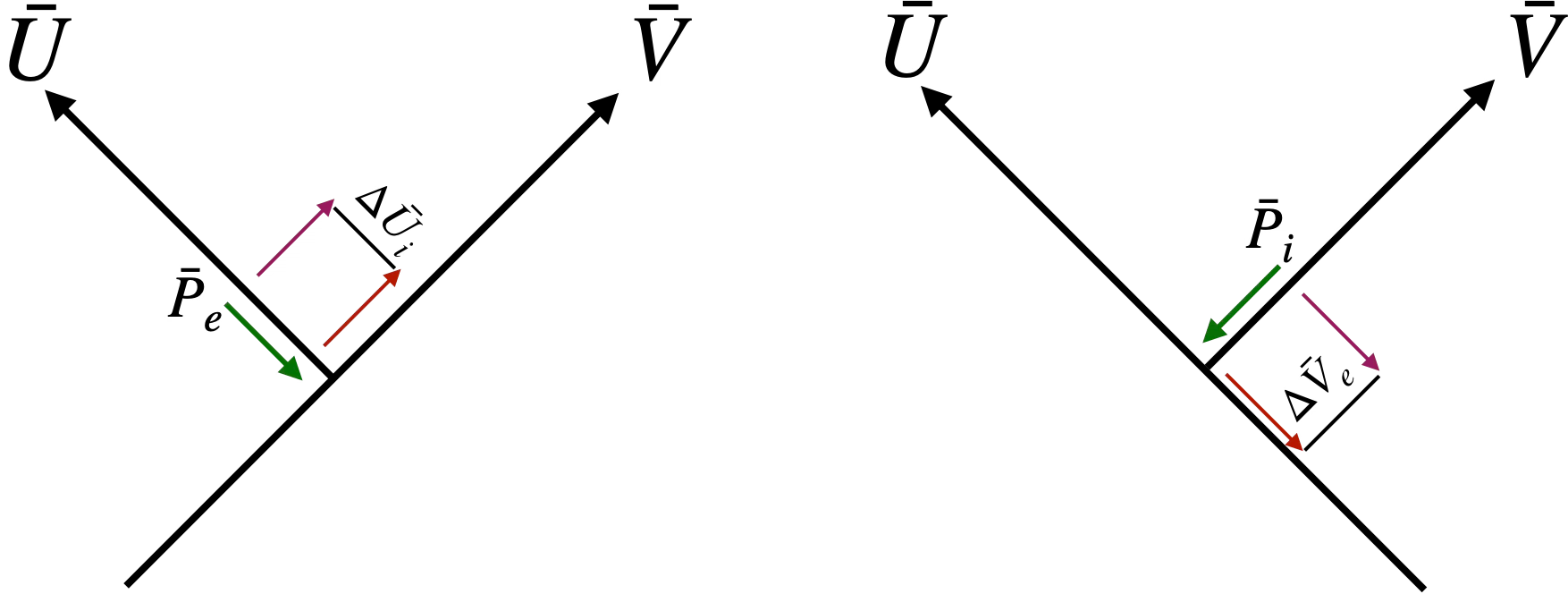}
	\caption{In the above picture, we depict the Dray and 'tHooft's gravitational backreaction effect associated with {i}ngoing and outgoing particles. A (classical) particle in the interior ($r\lesssim 2GM$, i.e., {$r< 2GM$ but we apply near horizon approximation}) with momentum $\bar P_i$ gravitationally cause a shift ($\Delta \bar V_e$) in the position of the exterior outgoing ($r\gtrsim 2GM$, i.e., {$r> 2GM$ but we apply near horizon approximation)} particle. Similarly, an exterior (classical) particle with momentum $\bar P_e$ gravitationally shifts ($\Delta U_i$) the interior classical particle position.   }
	\label{fig:gbr}
\end{figure}

Another assumption in the Hawking calculation is the commutation between the ingoing and outgoing quantum states. Following Eq. (2.4) of \cite{Hawking:1975vcx} the radial component of KG field operator\footnote{Defined by expansion in terms of spherical harmonics
\begin{equation}
	\phi \LF U,\,V \RF = \sum_{\ell,\, m} \frac{\Phi(U,\,V)}{r}Y_{\ell m}\LF \theta,\,\varphi \RF
\end{equation}} is expanded as the sum of interior and exterior parts as
\begin{equation}
	\hat \Phi = \hat \Phi_{in}+ \hat \Phi_{out} = \sum_i \LF p_i \textbf{b}_i + p_{i}^\ast \textbf{b}_i^\dagger\RF  +  \LF q_i \textbf{c}_i + q_{i}^\ast \textbf{c}_i^\dagger\RF
	\label{eq:Hexp}
\end{equation}
where $p_i$ describe purely outgoing states ($r>2GM$) and $q_i$ describe purely ingoing states, that end up inside $r<2GM$. In \cite{Hawking:1975vcx}, the following commutation relations are assumed 
\begin{equation}
\LT \hat \Phi_{in},\, \hat \Phi_{out} \RT =0\implies \LT \textbf{b}_i,\,\textbf{c}_i \RT =\LT \textbf{b}_i,\,\textbf{c}^\dagger_i \RT=0\,,
\label{eq:cominout}
\end{equation}
The points worth to be paid attention here are: 
\begin{itemize}
	\item The commutation relation \eqref{eq:cominout} is driven by the intuition that an interior state cannot affect the exterior state. 
	\item The expansion \eqref{eq:Hexp} treats the interior and exterior states as part of the same Hilbert space, with the same notion of time (in the QFT language), therefore ingoing and outgoing quanta belong to the same Fock space where time is uniquely defined.
\end{itemize}
 There are caveats in the above assumptions 
\begin{itemize}
	\item Following Dray and 't Hooft's gravitational backreaction \cite{Dray:1984ha}, an ingoing particle in Schwarzschild spacetime moving along the direction $V$ with momentum $P_i$ creates a shock wave (transverse field which is a function of $\theta,\, \varphi$) that causes a shift in the position of the outgoing particle that is approaching in the $U$ direction. The position of the outgoing particle ($V_e$) is affected by ingoing momenta ($P_i$), and the position of the ingoing particle ($U_i$) is affected by outgoing momenta ($P_e$) in the following way\footnote{We note that the 'tHooft considerations ignore interior region of black hole spacetime \cite{tHooft:2016rrl,tHooft:2022bgo}. In our approach, we consider both interior and exterior regions of SBH in the near horizon approximation (as schematically depicted in Fig.~\ref{fig:gbr}), and we take into account the gravitational backreaction effects between interior and exterior quanta \cite{Kumar:2023hbj} to define our direct-sum QFT that leads to unitarity. Thus, our construction is technically and conceptually very different from that of 't Hooft \cite{tHooft:2022umh} and also the works followed by other authors \cite{Betzios:2016yaq,Betzios:2020wcv,Betzios:2020xuj,Gaddam:2020mwe,Gaddam:2020rxb}.} \cite{Betzios:2016yaq,tHooft:2015pce,tHooft:2022umh} (See \cite{Kumar:2023hbj} for more details)
	\begin{equation}
		U_i = -\frac{8\pi G}{ R^2\LF \ell^2+\ell +1 \RF}P_e,\quad V_e = \frac{8\pi G}{R^2 \LF \ell^2+\ell +1 \RF}P_i\,. 
		\label{inoeff}
	\end{equation}
	where the subscripts $P_i,\,P_e$ and $V_e,\,U_i$ represent (dimensionless) interior and exterior (or ingoing and outgoing) momentum and positions (after the gravitational shift) that are defined by spherical harmonics (See Eq.~(5.1) of \cite{tHooft:2022umh}) 
\begin{equation}
	\begin{aligned}
V_e & \to \sum_{\ell,\, m} V_e Y_{\ell m}\LF \theta,\,\phi \RF,\quad U_i \to \sum_{\ell,\, m} U_i Y_{\ell m}\LF \theta,\,\phi \RF \\
		P_e & \to \sum_{\ell,\, m} P_e Y_{\ell m}\LF \theta,\,\phi \RF,\quad P_i \to \sum_{\ell m} P_i Y_{\ell m}\LF \theta,\,\phi \RF
	\end{aligned}
 \label{grb}
\end{equation}
It is important to note that $U_i$ is the position of the ingoing state (or the interior state at $r\lesssim 2GM$) defined for the KS coordinates $U<0,\,V>0$ where $V_e$ is the position of the outgoing state (or the exterior state at $r\gtrsim 2GM$)  for the KS coordinates $U>0,\,V>0$. Similarly, $V_i,\, U_e$ can be equally defined for the choice of KS coordinates $U>0,\,V<0$ and $U<0,\,V>0$. 

 One can now apply position and momentum uncertainty relations to the operators of interior and exterior states as \cite{tHooft:2022umh}
 \begin{equation}
     \LT \hat V_e,\,\hat P_e\RT = \LT \hat U_i,\,\hat P_i \RT = i 
     \label{canoBH}
 \end{equation}
Substituting \eqref{inoeff} in \eqref{canoBH} yields us
 \begin{equation}
     \LT \hat U_i,\,\hat V_e  \RT = i\frac{8\pi G}{R^2\LF\ell^2+\ell+1\RF}
     \label{thooftb}
 \end{equation}
\end{itemize}
Note that \eqref{thooftb} is the result of combining gravitational backreaction \eqref{inoeff} between the interior and exterior states (See Fig.~\ref{fig:gbr}) and the canonical non-commutative relations \eqref{canoBH}. This is the first quantization (i.e., description of a single quantum state in Schwarzshild spacetime) together with consideration of gravitational backreaction. For the second quantization\footnote{Note that we uplift position and momentum non-commutative relations of first quantization to the field and its conjugate momenta operators in the second quantization. On top of this, we demand all the operators commute for space-like distances, which is the step where we implement the relativistic principle. }, i.e., the description of quantum fields in Schwarzschild spacetime,  
we uplift the above relations and apply them to the interior and exterior field operators as \cite{Kumar:2023hbj}
\begin{equation}
    \LT \hat \Phi_{in},\, \hat \Phi_{out} \RT =  i\frac{8\pi G}{R^2\LF\ell^2+\ell+1\RF}\,, 
    \label{eq:qftalg}
\end{equation}
where $R=2GM$. Thus \eqref{eq:cominout} is valid only if we ignore the effects of gravitational backreaction or consider an infinitely large BH i.e., $R\to \infty$. In the next section, we discuss the construction of direct-sum QFTCS, which takes into account the discrete symmetries of SBH metric \eqref{dssym} and the result of gravitational backreaction algebra \eqref{eq:qftalg}. We emphasize that our consideration of gravitational backreaction effect is between interior and exterior state, which is different from 't Hooft \cite{tHooft:2016qoo,tHooft:2016rrl}. Note that the exterior $r\gtrsim 2GM $ regions in our picture are parity conjugate regions with opposite arrows of time (See Fig.~\ref{fig:confdia}). Furthermore, there cannot be any gravitational backreaction $I_{ext},\,II_{ext}$ regions in our picture because the points $\LF\theta,\,\varphi\RF$ and $\LF \pi-\theta,\,\pi+\varphi \RF$ are causally separated points for $r\gtrsim 2GM$. Also, our construction is different from \cite{Gaddam:2020mwe}, where authors do not take into account the quantization of the field in the interior and exterior regions of Schwarzschild spacetime. 

\section{Direct-sum QFT in curved spacetime and Hawking radiation with pure states}
\label{sec:disumBH}

\begin{figure}
    \centering
    \includegraphics[width=0.8\linewidth]{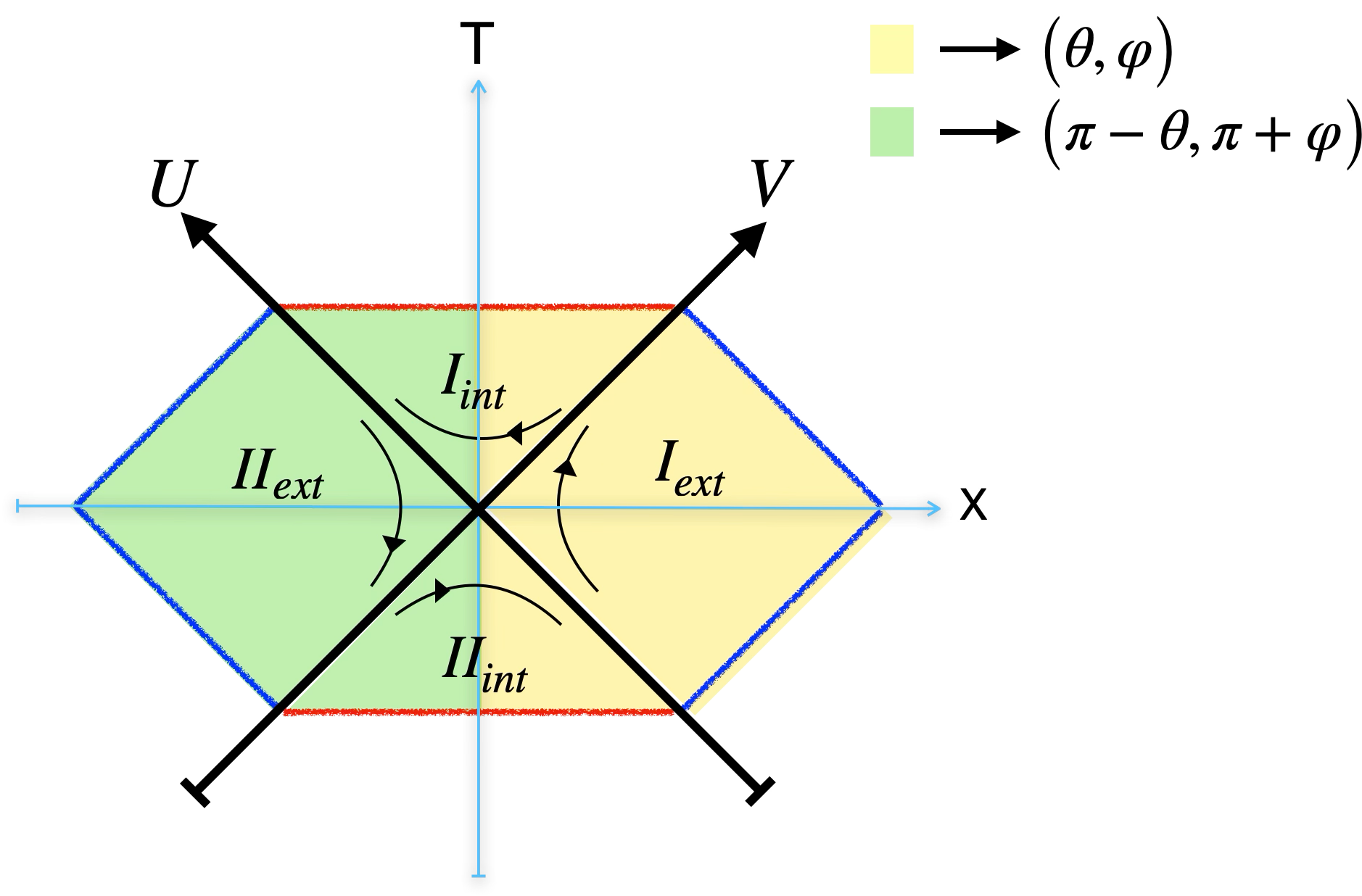}
    \caption{This is conformal diagram of direct-sum QFTCS in SBH spacetime \eqref{UVmetric} in the Kruskal coordinates $\LF U,\,V \RF$. The yellow and green shared regions indicate parity conjugate points $\LF\theta,\,\varphi\RF$ and $\LF\pi-\theta,\,\pi+\varphi\RF$ respectively. The regions labeled $I_{ext},\,II_{ext}$ represent $r>2GM$ where quantum field components lead to states evolving forward ($T: -\infty\to \infty$) and backward ($T:\infty\to -\infty$) in time $T=\frac{U+V}{2}$ whereas $I_{int},\,II_{int}$ represent interior $r<2GM$ where quantum field components lead to the states with opposite time evolutions. The red lines indicate the Schwarzchild singularity at $r=0$, which are identified in the regions $I_{int}$ and $II_{int}$. This conformal diagram does not contain any white hole or parallel Universe. All the regions are physical; they quantum mechanically indicate the evolution of quantum fields in SBH spacetime.}
    \label{fig:confdia}
\end{figure}

Direct-sum quantum theory, presented in Sec.~\ref{sec:disumQM}, forms a new basis for understanding quantum fields in SBH spacetime. With the direct-sum QFT in Minkowski spacetime, we can address the issue of two arrows of time associated with $U<0,\, V>0$ and $U>0,\,V<0$ of SBH metric \eqref{UVmetric} and the asymptotic limit $r\to \infty$. A simple thumb rule of direct-sum QFT is to form geometric superselection sectors for the Hilbert space (or Fock space) subjected to all the discrete spacetime transformations.  In the context of SBH, the metric \eqref{UVmetric} and the discrete symmetries \eqref{dssym} form guidelines for the description of quantum fields in the sectorial Fock spaces represented by the regions in the conformal diagram in Fig.~\ref{fig:confdia}. 

In our framework, we expand the field operator as direct-sum of interior ("int") and exterior ("ext") components as  \begin{equation}
	\begin{aligned}
		\hat{\Phi}  & = \hat{\Phi}_{int} \oplus \hat{	\Phi}_{ext} \\ 
		& = \frac{1}{\sqrt{2}} \LF  \hat{\Phi}^{I}_{int} \oplus \hat{	\Phi}^{II}_{int} \RF \oplus \frac{1}{\sqrt{2}}\LF  \hat{\Phi}^{I}_{ext} \oplus \hat{	\Phi}^{II}_{ext} \RF\, 
	\end{aligned}
 \label{eq:extintf}
\end{equation}
which are defined in the total Fock space, which is a direct-sum of geometric superselection sectors corresponding to the different spatial regions and notions of time ($r\lesssim 2GM$ and $r\gtrsim 2GM$) (See Fig.~\ref{fig:confdia})
\begin{equation}
	\Fc_{BH} = \LF\Fc_{I int}\oplus \Fc_{II int}\RF \oplus  \LF\Fc_{I ext} \oplus \Fc_{II ext} \RF
\end{equation}
For an asymptotic observer at $r\to \infty$ ($r_\ast \to \infty$) of SBH space-time \eqref{SBHmet}, the quantum field is in Minkowski space-time and it becomes
\begin{equation}
	\hat{\Phi}_{ext} = \frac{1}{\sqrt{2}}\LF \hat{\Phi}_{I \infty} \oplus \hat{\Phi}_{II \infty} \RF\,, 
 \label{infphi}
\end{equation}
corresponding to an asymptotic vacuum 
\begin{equation}
	\vert 0\rangle_{\infty} = \vert 0\rangle_{I \infty} \oplus \vert 0\rangle_{II \infty}, 
\end{equation}
defined by the annihilation operators 
\begin{equation}
\hat{a}_{I \tilde{\omega}}  \vert 0\rangle_{I\infty} =0,\quad \hat{a}_{II\, \tilde{\omega}} \vert 0\rangle_{II \infty} =0 \,. 
\end{equation}
The vacuum for quantum field $\hat{	\Phi}_{ext}$ in the near horizon approximation $r\to 2GM$ is  
\begin{equation}
    \vert 0\rangle_{ext} = 
      \vert 0\rangle_{Iext}   \oplus   \vert 0\rangle_{IIext}
\end{equation}
defined by the annihilation operators 
\begin{equation}
\hat{b}_{I {\omega}}  \vert 0\rangle_{I ext} =0,\quad \hat{b}_{II\, {\omega}} \vert 0\rangle_{II ext} =0 \,. 
\end{equation}

By applying the technique of Bogoliubov transformation, we obtain the number density of particles created by operators $\hat{a}^\dagger_{I\,k},\, \hat{a}^\dagger_{II\,k}$ in the vacuum $\vert 0\rangle_{BH}$ which is what exactly asymptotic observer see as Hawking particles. Following \cite{Mukhanov:2007zz} we can compute
\begin{equation}
	\begin{aligned}
	a_{I \tilde{\omega}}  & = \int_{0}^{\infty} d\omega \LF \alpha_{\omega\tilde{\omega}} b_{I\,\omega}  +\beta_{\omega\tilde{\omega}} b^\dagger_{I\,\omega} \RF \\
		a_{II \tilde{\omega}}  & = \int_{0}^{\infty} d\omega \LF \tilde{\alpha}_{\omega\tilde{\omega}} b_{II\,\omega}  +\tilde{\beta}_{\omega\tilde{\omega}} b^\dagger_{II\,\omega} \RF
	\end{aligned}
\end{equation}
where
\begin{equation}
	\begin{aligned}
		\alpha_{\omega\tilde{\omega}} & = \sqrt{\frac{\tilde{\omega}}{\omega}} \int_0^\infty  \frac{du}{2\pi} e^{i\tilde{\omega} u-i\tilde{\omega} U}, \quad 	\beta_{\omega\tilde{\omega}} & = \sqrt{\frac{\tilde{\omega}}{\omega}}\int_0^\infty  \frac{du}{2\pi} e^{i\tilde{\omega} u+i\tilde{\omega} U} \\ 
		\tilde{	\alpha}_{\omega\tilde{\omega}} & = \sqrt{\frac{\tilde{\omega}}{\omega}}\int_0^\infty  \frac{du}{2\pi} e^{-i\tilde{\omega} u+i\tilde{\omega} U}, \quad 	\tilde{	\beta}_{\omega\tilde{\omega}} & = \sqrt{\frac{\tilde{\omega}}{\omega}}\int_0^\infty  \frac{du}{2\pi} e^{-i\tilde{\omega} u-i\tilde{\omega} U}
	\end{aligned}
\end{equation}
Thus, the asymptotic observer witness Hawking radiation by the follow number of quanta 
\begin{equation}
	\begin{aligned}
 N_{\tilde{\omega}} & = \frac{1}{2}	{}_{BH}\langle 0 \vert \LF \hat{a}_{I\,\tilde{\omega}}^\dagger \oplus \hat{a}_{II\,\tilde{\omega}}^\dagger  \RF \LF \hat{a}_{I\,\tilde{\omega}} \oplus \hat{a}_{II\,\tilde{\omega}} \RF  \vert 0\rangle_{BH}  \\ 
& = \frac{1}{2} \int_{0}^{\infty}d\omega \LF \vert \beta_{\omega\tilde{\omega}} \vert^2+ \vert \tilde{	\beta}_{\omega\tilde{\omega}} \vert^2 \RF\, \\ 
& = \frac{1}{e^{8\pi GM\tilde \omega/\hbar}-1}\,,
\end{aligned}
\end{equation} 
which is thermal distribution with a temperature $T=\frac{\hbar}{8\pi GM}$. But here, in our context, Hawking radiation is in the form of pure states. Because of the direct-sum structure of exterior and interior field operators \eqref{eq:extintf}, which is different from the one of Hawking's consideration \eqref{eq:Hexp}, any maximally entangled non-separable pure state becomes a direct-sum of two pure state components 
\begin{equation}
\begin{aligned}
    \vert \Tilde{\psi}_{12}\rangle =\sum_{m,n} \Tilde{c}_{mn} \vert \Tilde{\phi}_{m1}\rangle  \otimes \vert \Tilde{\phi}_{n2}\rangle & = \frac{1}{\sqrt{2}}\sum_{m,n} \Tilde{c}_{mn} \LF \vert \Tilde{\phi}^{\rm ext}_{m1}\rangle  \otimes \Tilde{\phi}^{\rm ext}_{n2}\rangle \RF \oplus \LF \vert \Tilde{\phi}^{\rm int}_{m1}\rangle  \otimes \Tilde{\phi}^{\rm int}_{n2}\rangle \RF \\ 
    & = \frac{1}{\sqrt{2}} \begin{pmatrix}
        \vert \Tilde\psi^{ext}_{12}\rangle  \\ \vert \Tilde\psi^{int}_{12}\rangle 
    \end{pmatrix}
    \end{aligned}
\end{equation}
where $\Tilde{c}_{mn}\neq \Tilde{c}_n\Tilde{c}_m$, $\vert \phi_1\rangle = \sum_m \tilde{c}_m\vert \phi_{m1}\rangle  $ and $\vert \phi_2\rangle = \sum_n \tilde{c}_n\vert \phi_{n2}\rangle  $. This would render any density matrix of pure states to become two direct-sum counterparts which are pure states on their own. 
\begin{equation}
    \rho = \frac{\rho_{ext}}{2}\oplus \frac{\rho_{int}}{2}.
\end{equation}
The Von Neumann entropies of exterior and interior (pure) state components ($S_{ext},\, S_{int}$) vanish since exterior and interior are geometric superselection sectors
\begin{equation}
    S_{ext} = -Tr\LF\rho_{ext}\ln \rho_{ext}\RF=0,\quad S_{int} = -Tr\LF\rho_{int}\ln \rho_{int}\RF=0,\quad S_{H} = S_{ext}+S_{int}=0\,,
\end{equation}

This means any asymptotic observer, irrespective of what is beyond the horizon, would witness the pure states evolving into pure states in his/her geometric superselection-sector of the total Hilbert space. Since states behind the horizon are related by discrete spacetime transformation $UV<0 \to UV>0$ (See region Fig.~\ref{fig:confdia}), one can reconstruct the interior state of SBH. Furthermore, because of non-commutative relation \eqref{eq:qftalg}, the exterior hawking quanta are not independent of the interior counterparts, which leads to information reconstruction. This forms the first step towards resolving the information paradox. 

\section{Conclusions and outlook}
\label{sec:con}

Even after decades of developments in quantum gravity research, the deepest problem in theoretical physics is at a scale much below the Planck scale, which is the unitarity problem at the gravitational horizons \cite{Giddings:2022jda}. Quantum field theory in curved spacetime requires much stronger foundations, which would help to improve our efforts toward building quantum gravity at the Planck scales. 

Drawing on foundational work by Hawking and 't Hooft, we propose a QFT framework that incorporates discrete spacetime transformations, acknowledging the status of time as a parameter (i.e., not an operator in quantum theory). This approach, tested in various spacetimes such as Rindler \cite{Kumar:2024oxf}, de Sitter, Minkowski, and inflationary universe \cite{Kumar:2023ctp,Gaztanaga:2024whs,Kumar:2022zff}. The framework has been successful in explaining the long-standing CMB anomalies, which are largely associated with parity asymmetry \cite{Gaztanaga:2024whs}. In this paper, we presented a discussion of how direct-sum QFTCS promises new insights into BH physics \cite{Kumar:2023hbj}.
The proposed quantization approach, incorporating 't Hooft's gravitational backreaction effects, has unveiled a novel framework for understanding quantum fields in black hole spacetime, specifically within the context of Schwarzschild black holes. In this review, we contemplated the idea that the black hole under the rules of direct-sum QFTCS allows us to redefine entanglement between interior and exterior Hawking quanta in the form of pure states. Concurrently, we observed that exterior radiation is intrinsically linked to the interior due to new non-commutative relations resulting from 't Hooft's gravitational backreaction algebra. This phenomenon, absent from Hawking's original 1975 paper, marks a significant departure from traditional understandings.

We have explored how direct-sum QFT provides a robust framework for maintaining unitarity and observer complementarity (by means of pure states evolving into pure states) in the presence of spacetime horizons. By decomposing the Hilbert space into a direct sum of sectorial Hilbert spaces, each associated with different geometric superselection sectors (linked to $\mathcal{P}\mathcal{T}$ symmetry), DQFT allows for a more nuanced understanding of information conservation.
This framework could allow information retrieval beyond the horizons by splitting Hilbert or Fock space into geometric superselection sectors.

In conclusion, the study of Hawking radiation with pure states remains a profound area of research within theoretical physics, particularly concerning the interplay between quantum mechanics and gravity. Hawking's initial proposal highlighted the intrinsic challenges, such as the information paradox, unitarity violation, and predictability breakdown, which arise from quantizing a scalar field in Schwarzschild black hole spacetime. These issues, traceable to early observations made by Einstein and Rosen on the fundamental conflict between gravity and quantum mechanics \cite{Einstein:1935tc,GKM}, continue to spur significant advancements in the field of quantum gravity. Direct-sum quantum theory in curved spacetime by reinstating the unitarity would initiate a new foundational approach towards quantum gravity. 

\section*{Acknowledgements}

KSK thanks the support of the Royal Society for the Newton International Fellowship. This research was funded by Funda\c{c}\~ao para a Ci\^encia e a Tecnologia grant number UIDB/MAT/00212/2020. KSK would like to thank the organizers of the EREP 2023 for the kind invitation and opportunity to give a talk on the subject related to this paper. Authors would like to thank E. Gaztanaga and Mathew Hull for the useful discussions. 

\bigskip









\bibliography{sn-bibliography}

\end{document}